\documentclass[prd,amsmath,amssymb,twocolumn]{revtex4-2}
\pdfoutput=1
\usepackage[utf8]{inputenc}
\usepackage{amsmath}
\usepackage{comment}
\usepackage{graphicx}
\usepackage{siunitx}
\usepackage{tikz}
\usepackage{hyperref}
\usetikzlibrary{positioning}
\usepackage{float}
\usepackage{subcaption}
\usepackage{appendix}
 \usepackage{mathtools}
 \usepackage{nccmath}

\newcommand\pasp{PASP}

\begin{document}

\title{Bayesian Time Delay Interferometry for Orbiting LISA: Accounting for the Time Dependence of Spacecraft Separations}

\author{Jessica Page}
\affiliation{Space Science Department, University of Alabama in Huntsville, 320 Sparkman Drive, Huntsville, AL 35899, USA}
\author{Tyson B. Littenberg}
\affiliation{NASA Marshall Space Flight Center, Huntsville, AL 35812, USA}

\begin{abstract}
Previous work demonstrated effective laser frequency noise (LFN) suppression for Laser Interferometer Space Antenna (LISA) data from raw phasemeter measurements using a Markov Chain Monte Carlo (MCMC) algorithm with fractional delay interpolation (FDI) techniques to estimate the spacecraft separation parameters required for time-delay interferometry (TDI) under the assumption of a rigidly rotating LISA configuration. Including TDI parameters in the LISA data model as part of a global fit analysis pipeline enables gravitational wave inferences to be marginalized over uncertainty in the spacecraft separations. Here we extend the algorithm's capability to perform data-driven TDI on LISA in Keplerian orbits, which introduce a time-dependence in the arm-length parameters and at least $\mathcal{O}$(M) times greater computational cost since the filter must be applied for every sample in the time series of sample size M. We find feasibility of arm-length estimation on $\sim$day-long time scales by using a novel Taylor-expanded version of the fractional delay interpolation filter that allows half of the filter computation to be calculated and stored before MCMC iterations and requires shorter filter lengths than previously reported. We demonstrate LFN suppression for orbiting LISA using accurate arm-length estimates parameterized by Keplerian orbital parameters under the assumption of unperturbed analytical Keplerian orbits, and explore the potential extension of these methods to arbitrary numerical orbits.

\end{abstract}

\maketitle

\section{Introduction}\label{introduction_paper_2}

Astrophysical sources of gravitational waves (GW) in the 0.1 mHz -- 0.1 Hz band will be detected through the space-based LISA (Laser Interferometer Space Antenna) mission set to launch in the mid 2030s. LISA combines three spacecraft separated by a mean inter-spacecraft distance of ${\sim}\SI{2.5e6}{km}$ to measure the pathlength change of the free-falling test masses induced by GWs using laser interferometry \cite{LISA}. 

Because LISA is in a rotating heliocentric orbit, the inter-spacecraft distances will differ and vary in time,  leaving a fractional frequency fluctuation amplitude of $\frac{\Delta \nu}{\nu} = \SI{e-13}{Hz^{-1/2}}$ remaining in the data. Typical GW signals are found near $\SI{e-21}{Hz^{-1/2}}$, requiring laser frequency noise (LFN) suppression. Time-delay interferometry (TDI) is a well-studied algebraic operation to suppress LFN adequately below other secondary noises such as the test-mass acceleration and optical metrology system noises for GW signal analysis \cite{TDI_review}. TDI works by linearly combining the LISA interferometric measurements delayed appropriately by the light travel time between spacecraft to synthesize a digital equal-arm interferometer. The LISA mission will use psuedo-random noise phase modulation (PRM) encoded in the measurements to obtain the spacecraft separations $L$ (i.e. ``arm-lengths'', delays $L/c$) needed in the TDI combinations.

There are alternative data-driven methods that provide a back-up method to PRM that suppress LFN from the raw LISA measurements such as \cite{secondGenAPCI, Baghi_et_al_2021,Paper_1,TDIR}. This paper is an extension of \cite{Paper_1} to estimate the time-varying delays needed for second-generation TDI that accounts for the heliocentric motion of the LISA array using a Markov chain Monte Carlo (MCMC) algorithm. The use of MCMC methods to estimate arm length posteriors allows one to marginalize over their uncertainty when computing the LISA response function to GWs and opens the possibility to integrate TDI into the LISA ``global fit'' as demonstrated in \cite{globalFit}. High signal to noise GW sources were found to require accurate orbital models that account for the arm length time-dependence when computing the GW response function \cite{fastlisaresponse}, indicating the importance of quantifying the spacecraft separation uncertainty. 

Bayesian TDI builds upon Time-delay Interferometric Ranging (TDIR) which estimates the arm lengths from raw interferometer measurements by minimizing the power of the residual LFN \cite{TDIR} which relies on fractional delay interpolation (FDI) \cite{FDI} to estimate the required nanosecond precision delays that occur between samples telemetered to Earth. Our previous work used an MCMC to estimate the delays in a rigidly rotating LISA scenario where the clock-wise and counter clockwise light travel times differ but remained time-independent. We found that a shorter filter length can be used by limiting the maximum frequency in the likelihood function to reduce data loss in the presence of data gaps. This work extends the ideas of Bayesian data-driven TDI to assess feasibility in the realistic case of LISA heliocentric motion. We review second-generation TDI and describe the relevant LISA measurement details in Sec. \ref{measurement-description}. We introduce a new highly efficient formulation of the LaGrange FDI filter in Sec. \ref{newFilterSection} and examine the computational optimization that would be crucial for Bayesian TDI methods. We explore the MCMC in Sec. \ref{MCMCSection}, and a way to parameterize the arm length time dependence in the case of LISA on Keplerian orbits in Sec. \ref{KeplerianParameterization}. Finally we demonstrate a promising method to estimate arm lengths in the realistic case of arbitrary non-Keplerian orbits that account for all other forces other than solar gravitational motion in Sec. \ref{arbitraryOrbits}.

\section{LISA Measurement Description and Second-Generation TDI}\label{measurement-description}

\begin{figure}
    \includegraphics[scale=0.7]{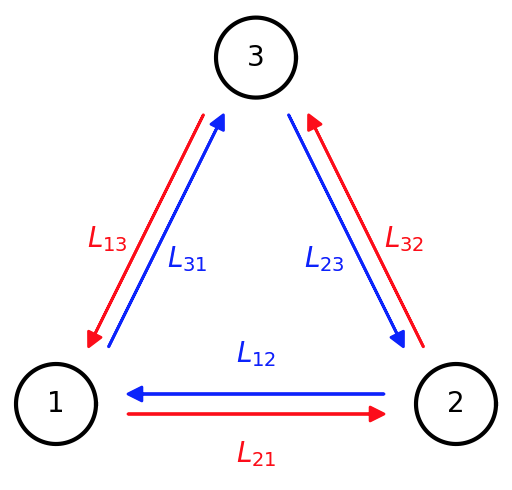}
    \caption{Arm length labeling convention in relation to spacecraft number. Figure does not reflect that each spacecraft contains two test masses and two optical measurement systems.}
    \label{fig:diagram}
\end{figure}

We use the LISA Simulation software \cite{LISAInstrumentArticle} for the data simulation to account for the heliocentric motion of the rotating array and therefore follow the LISA labeling conventions used there. The arm length labels in reference to each spacecraft are demonstrated in \ref{fig:diagram}. The meaning of the first and second indices of a measurement has essentially been alternated from the notation in \cite{Paper_1}, and we again follow the split interferometry design described in \cite{otto_2012}. The science measurement $s_{ij}$ contains the GW signal and is the measurement exchanged between two separate spacecraft; $s_{ij}$ is received on spacecraft $i$ from $j$. The test mass measurement $\varepsilon_{ij}$ is the motion of the test mass on spacecraft $i$ that is adjacent to spacecraft $j$, and the reference measurement on $\tau_{ij}$ interferes the light between the two optical benches on a given spacecraft and the $i$ indice is denoted as the receiving. The distance between two spacecraft will differ depending on the direction the laser light is traveling due to the Sagnac effect ($L_{ij} \neq L_{ji}$). The $i$ indice in the arm lengths denotes the receiving spacecraft and the $j$ indice denotes the emitting spacecraft. We ignore all secondary noises and GW in the simulated data (sec. \ref{data}) for this demonstration but we show the laser phase noise $p_{ij}$ on the optical bench on spacecraft $i$ that is adjacent to spacecraft $j$ that enters in the three interferometric measurements given in Eqns. \ref{eq:measurments1} and \ref{eq:measurements2} and include our assumption of the test-mass acceleration noise $n^{\textrm{TM}}_{ij}$ and optical metrology system noise $n^{\textrm{OMS}}_{21}$ placement. Cyclic permutation of both indices for both sets of Eqns. \ref{eq:measurments1} and \ref{eq:measurements2} yield the remaining 12 measurements.

\begin{align}
    s_{21} &= p_{12,21}-p_{21} + n^{\textrm{OMS}}_{21} \nonumber \\
    \tau_{21} &= p_{23}-p_{21} \nonumber \\
    \varepsilon_{21} &= p_{23}-p_{21} - 2 n^{\textrm{TM}}_{21}\label{eq:measurments1} \\
    s_{23} &= p_{32,23}-p_{23} + n^{\textrm{OMS}}_{23} \nonumber \\
    \tau_{23} &= p_{21}-p_{23} \nonumber \\ 
    \varepsilon_{23} &= p_{21}-p_{23} + 2 n^{\textrm{TM}}_{23}\label{eq:measurements2}
\end{align}

\subsection{TDI 2.0 for LISA in Heliocentric Orbit}\label{TDI20}

The main purpose of this work is to demonstrate feasibility of Bayesian-estimated arm-lengths for the realistic orbiting LISA scenario. The heliocentric motion of the three spacecraft array introduces a time dependence on $L_{ij}(t)$ with $\dot{L}_{ij} \sim 10$ m/s which leaves fractional frequency fluctuations present on $\mathcal{O}(10)$ times above the secondary noises in the GW signal-rich portion of the LISA band \cite{reference_TDIR,TDI_review} if the ``first generation'' combinations in the previous Bayesian TDI work \cite{Paper_1} are used. The ``second-generation'' (i.e TDI 2.0) combinations average out the time-dependent effects by linearly approximating each light path ($L_{ij}(t) \approx L_{ij} + \dot{L}_{ij}t$) and synthesizing two equal light paths that now strongly depend on the order that each one-way beam is applied \cite{first_paper_second_gen} which now requires non-commuting delay operators $\mathcal{D}_{ij}$ in the second-generation combinations. LFN is suppressed sufficiently to first order in spacecraft distance velocities $\dot{L}_{ij}$. We continue use of the common 16-link Michaelson combinations (Eq. \ref{eq:TDI_20}) denoted with a ``2'' subscript ($X_{2}(t)$, $Y_{2}(t)$, $Z_{2}(t)$) since they are second-generation combinations. The $_{;\mathcal{D}_{ij}\mathcal{D}_{mn}}$ notation indicates the time delay operator $\mathcal{D}_{mn}\mathcal{D}_{ij}$ that acts on a function $f(t)$ as given in Eq. \ref{eq:delayOperator}, and similarly for more delays; the operator closest to a measurement is applied first and subsequent delays act on the measurement and previously-applied delays. The use of a semi-colon indicates the delays do not commute to keep standards in the literature. The first order expansion to the arm length rate of change is shown in Eq. \ref{eq:LineardelayOperator}, and the comma $_{,\mathcal{D}_{ij}\mathcal{D}_{mn}}$ notation denotes constant valued estimates of commuting operators.

\begin{alignat}{2}
    \mathcal{D}_{mn}\mathcal{D}_{ij}f(t) &= f(t)_{;\mathcal{D}_{ij}\mathcal{D}_{mn}}\\
    &= f(t-\frac{L_{ij}(t-L_{mn}(t)/c)}{c})\label{eq:delayOperator} \\
    &\approx f_{,\mathcal{D}_{ij}\mathcal{D}_{mn}} + \dot{f}_{,\mathcal{D}_{ij}\mathcal{D}_{mn}}\dot{L}_{ij}L_{mn}
\label{eq:LineardelayOperator}
\end{alignat}

Cyclic permutation of both indices of Eq. \ref{eq:TDI_20} yield the $Y_{2}(t)$ and $Z_{2}(t)$ combinations for the light paths of the other two spacecraft.

\begin{multline}
    X_2(t) = [\mathcal{D}_{12}\mathcal{D}_{21}\mathcal{D}_{13}\mathcal{D}_{31} - \mathcal{I}] [(\eta_{13} + \eta_{31;\mathcal{D}_{13}})\\
    + (\eta_{12}+\eta_{21;\mathcal{D}_{12}})_{;\mathcal{D}_{31}\mathcal{D}_{13}}]\\
    - [\mathcal{D}_{13}\mathcal{D}_{31}\mathcal{D}_{12}\mathcal{D}_{21} - \mathcal{I}] [(\eta_{12} + \eta_{21;\mathcal{D}_{12}})\\
    + (\eta_{13}+\eta_{31;\mathcal{D}_{13}})_{;\mathcal{D}_{21}\mathcal{D}_{12}}]
    \label{eq:TDI_20}
\end{multline}

The intermediary variables $\eta_{ij}$ in Eq. \ref{eq:TDI_20} are combinations of the interferometric measurements that remove 3 of the 6 LFN terms, essentially making the LFN the same on both optical benches on a given spacecraft. Cyclic permutation of both indices twice for both equations of Eq. \ref{eq:intermediary} provide the 6 intermediary variables.

\begin{alignat}{2}
    \eta_{12} &= s_{12} +\frac{\tau_{12}-\varepsilon_{12}}{2} + D_{12}\frac{(\tau_{21}-\varepsilon_{21})}{2} - D_{12}\frac{\tau_{21}-\tau_{23}}{2}\nonumber  \\
    \eta_{13} &= s_{13} +\frac{\tau_{13}-\varepsilon_{13}}{2} + D_{13}\frac{(\tau_{31}-\varepsilon_{31})}{2} + \frac{\tau_{12}-\tau_{13}}{2}
\label{eq:intermediary}
\end{alignat}

\section{Bayesian Estimation of Time-Dependent Spacecraft Separations}\label{MCMCSection}

\subsection{MCMC}\label{mcmcSpecs}

The MCMC algorithm utilizes Bayes' Theorem to estimate the posterior distribution $p(\vec{\pmb{\theta}}|\vec{\mathbf{d}})$ of the delay parameters $\vec{\pmb{\theta}} = \vec{\mathcal{D}}_{ij}(t)$ given the data $\vec{\mathbf{d}} = \{\tilde{X}_2(f),\tilde{Y}_2(f),\tilde{Z}_2(f)\}$ which are the Fourier-transformed TDI combinations, and various parameterizations for the $L_{ij}(t)$ time dependence are described in Sec. \ref{KeplerianParameterization}. The FDI filter described in Sec. \ref{newFilterSection} is applied at each iteration of the chain to precisely determine the measurement time series delayed by the proposed arm-length. The prior distribution is uniform in all data models tested, but the parameters and parameter ranges vary depending on the data as described in Sec. \ref{results}.

The log likelihood function neglecting constant normalization factors is given in  Eq. \ref{eq:likelihood}. 

\begin{equation}\label{eq:likelihood}
    \ln p(\vec{\mathbf{d}}|\vec{\pmb{\theta}}) = \sum_{i=f_\textrm{min}}^{f_{\textrm{max}}} \left[-\ln(|\mathbf{C}|)_{i}-\left(\sum\limits_{j,k}^{X,Y,Z}\mathbf{d}_j^{\dagger}\mathbf{C}^{-1}_{jk}\mathbf{d}_k\right)_i\right],
\end{equation}

where $f_\textrm{min} = 10^{-4}\ {{\rm Hz}}$ and $f_\textrm{max}$ is restricted to 0.03 Hz instead of 1 Hz which is typically the assumed maximum of the LISA band. Interpolation error increases dramatically at frequencies near the 1 Hz maximum and we previously found in \cite{Paper_1} that a 0.1 Hz restriction (just below the transfer frequency $f_{*} = c/L$) yields better parameter accuracy using a lower sampling rate and significantly shorter filter length. We restrict the likelihood calculation to below 0.03 Hz to reduce the computational cost of time-varying FDI filters and examine LFN suppression to the most relevant parts of the LISA band. Assuming that LFN is white noise in the LISA band $10^{-4}-1$ Hz with fractional frequency power spectral density (PSD) $S^{\rm LFN}_{y} = 10^{-26}{\mathrm{Hz}^{-1}}$, contributions to the likelihood integral above 0.03 Hz are negligible.

The noise covariance matrix $\mathbf{C}(f)$ describes the TM and OMS noises that remain after TDI application in the $\mathbf{d}_j^{\dagger}$ and $\mathbf{d}_k$ terms. We assume equal and constant arm lengths in the calculation of $\mathbf{C}(f)$ and use the same $\mathbf{C}(f)$ that is applied to the rigidly rotating LISA scenario so that all diagonal elements are given by Eq. \ref{eq:diagonalCov},

\begin{multline}\label{eq:diagonalCov}
\mathrm{C}_{jj} = 16\sin^2{(2\pi f L)} S^{\mathrm{OMS}}_y\\
+(8\sin^2{(4\pi f L)}+32\sin^2{(2\pi f L)}) S^{\mathrm{OMS}}_y
\end{multline}

\noindent and all off-diagonal elements are given by Eq. \ref{eq:offdiagonalCov}

\begin{equation}\label{eq:offdiagonalCov}
\mathrm{C}_{jk} = (4S^{\mathrm{Tm}}_y+S^{\mathrm{OMS}}_y) (-4 \sin{(2\pi f L)} \sin{(4\pi f L)})
\end{equation}

\noindent since we are focused on arm length estimation in the case of data containing LFN-only. We found no effect on arm length estimation when the delay parameters were included in the unequal-arm noise covariance matrix in \cite{Paper_1}, but we could similarly derive the second-generation noise covariance matrix elements in future studies that include TM and OMS noises in the data.

We assume the TM and OMS noise PSDs converted to fractional frequency are as given in \cite{LISA} shown in Eq. \ref{eq:SecondaryNoisePSDs}.

\begin{widetext}
\begin{align}
    S^{\textrm{TM}}_{y}(f) &= (2\pi f c)^{-2} (\num{3e-15})^2 \left(1+\left(\frac{\num{4e-4}}{f}\right)^2\right)  \left(1+\left(\frac{f}{\num{8e-3}}\right)^4\right)  \textrm{Hz}^{-1}\nonumber\\
    S^{\textrm{OMS}}_{y}(f) &= \left(\frac{2 \pi f}{c}\right)^{2} (\num{1.5e-11})^2 \left(1+\left(\frac{\num{2e-3}}{f}\right)^4\right) \textrm{Hz}^{-1}.
    \label{eq:SecondaryNoisePSDs}
\end{align}
\end{widetext}

\subsection{Optimized LaGrange Fractional Delay Interpolation Filter}\label{newFilterSection}

This work extends upon the ideas of Bayesian data-driven TDI (i.e. time-delay interferometric ranging (TDIR) \cite{Paper_1} to include time-varying delays. TDIR was first introduced in \cite{TDIR} and relies heavily on the ideas of \cite{FDI} which showed that by interpolating between integer samples to land on delay estimates at nanosecond precision and minimizing the power of the residual LFN, one can perform TDI from the raw interferometric measurements and obtain LFN suppression sufficiently below the secondary noises. We previously used the same LaGrange-windowed fractional delay interpolation (FDI) filter as shown in \cite{FDI} and adjusted the filter length requirements for the sampling rate and cut-off frequency for the data simulated in \cite{Paper_1}. Now that the delays are time-varying, re-use of the same LaGrange filter would require a new $N^{\text{th}}$ order filter for every $M^{\text{th}}$ sample where $M$ is the number of samples in the data. The previously used LaGrange filter is too computationally costly for updating the filter each sample when working with stochastic sampling algorithms such as an MCMC, so we adapted the LaGrange-windowed sinc filter based off of \cite{new_Lagrange} to remove half of the computation out of every MCMC iteration which dramatically reduces total cost and is described as follows. Equation \ref{eq:FDI_before} is the FDI process for a constant delay that is applied to every sample of the time series.

\begin{align}
    f(n-D) &= f(n) \ast w_{L}(D)\mathrm{sinc}(n-D)\nonumber \\
    &= \sum_{k=-(N-1)/2}^{(N-1)/2} f(n+k)w_{L}(k,D)\mathrm{sinc}(D-k)
    \label{eq:FDI_before}
\end{align}

The LaGrange window is given in equation (7) of \cite{FDI}. We used the convolution theorem and computed the Fourier transforms of the LISA measurements before beginning MCMC iterations which significantly reduced the computation time in the constant delay case. Now, the overall delay applied to a measurement is a function of time so the interpolation filter is different for each sample in the time series (Eq. \ref{eq:FDI_after}). 

\begin{equation}
    f(n-D_{n}) = f(n) \ast w_{L}(D_{n})\mathrm{sinc}(n-D_{n})
    \label{eq:FDI_after}
\end{equation}

We implement a Taylor series expansion of the delayed signal based on \cite{new_Lagrange} in  terms of the difference operator $$\Delta f[n] = f[n]-f[n-1],$$ powers of the LISA measurements $f[n]$ and factorial polynomials of the delays $$D^{[k]} = D (D+1) \ldots (D+k-1)$$ to result in Eq. \ref{eq:newfiltersum}.

\begin{equation}\label{eq:newfiltersum}
    f(n-D_n) = \sum_{k=0}^{N} \Delta^{[k]} f[n] \frac{(-D_n)^{[k]}}{k!}
\end{equation}

The $\Delta^{[k]}f[n]$ powers of the measurements are computed and stored for a given $N^{\text{th}}$ order filter before running the MCMC iterations and $\Delta^{[k]}f[n] = \Delta\Delta^{[k-1]}f[n]$. Paper \cite{new_Lagrange} quotes $3N-2$ additions and $3N-1$ multiplications for an $N^{\text{th}}$ order filter per output sample. Storing the $\Delta^{[k]}f[n]$ factors of Eq. \ref{eq:newfiltersum} beforehand further reduces the new filter's additions to $N$ per output sample. This results in approximately 30 times greater computational speed improvement for one MCMC iteration on one hour of data. Following standard guidelines set by the LISA Consortium for interpolating filters, the filter is shifted so that the $n^{\text{th}}$ sample to delay is centered in the middle of the filter and both even and odd filter lengths can be used.

This new LaGrange filter formulation also requires shorter filter lengths than the standard LaGrange-windowed sinc or LaGrange polynomials. The previous Bayesian TDIR work that used Eq. \ref{eq:FDI_before} found that the minimum filter length N required was $\approx$ 23 for a sampling rate $f_s = 4$ Hz and a likelihood function maximum frequency of 0.1 Hz. The new Taylor series representation requires $N=7$ for the same sampling specifications resulting in 4 seconds of data saved each time the delay estimation process must be recomputed to account for gaps in mission data. 

Additionally, the code implements an algebraically-reduced and optimized version of the TDI combinations by substituting Eq. \ref{eq:intermediary} into  Eq. \ref{eq:TDI_20} and combining terms with the same delay operator combination so that each unique delay sequence need only be applied once which significantly reduces run-time. We implement the $\dot{f}_{,\mathcal{D}_{ij}}$ term in Eq. \ref{eq:LineardelayOperator} by taking the NumPy numerical gradient of the delayed signal with second order central differences of the interior points and first order forward and backwards differences for the end samples.

\subsection{Data}\label{data}

We use the \verb|LISA Instrument| software to simulate the interferometric measurements $s_{ij}$, $\tau_{ij}$ and $\varepsilon_{ij}$. They are given in units of fractional frequency fluctuations for a given sampling rate, data duration, data content, reference frame and orbit model using \verb|LISA Orbits|. We assume a 4 Hz sampling rate with varying duration depending on the orbit model used. We are demonstrating feasibility of Bayesian TDIR for second-generation TDI, so we consider LFN only for now which excludes the optical metrology system, test mass, clock and all other optional noises as well as GWs. We assume no up-sampled on-board physics and therefore no anti-aliasing filter, and keep the remaining arguments to \verb|LISA Instrument| at their default settings. Since we are interested in examining Bayesian estimates of time-varying arm lengths, we explore both the Keplerian and the numerically computed ESA (European Space Agency) orbit models in \verb|LISA Orbits| as input to \verb|LISA Instrument|. The \verb|Keplerian| orbit model assumes the center of mass of the LISA constellation is on a heliocentric orbit. A more realistic numerical orbit model \verb|ESA Orbits| accounts for gravitational forces from the Earth and other planetary bodies, as well as other non-gravitational forces \cite{Trajectory_Design}. The \verb|LISA Instrument| output is in units of frequency fluctuations $\SI{}{Hz}$ and we convert to fractional frequency fluctuations $\SI{}{Hz Hz^{-1}}$ so we implement the suggestion given in \cite{Bayle_Staab_2021} to mitigate for the Doppler shift caused by the armlength variation in Eq. \ref{eq:TDI_20} by using the Doppler delay operator in place of delay operators.

\begin{figure*}
\begin{subfigure}{.5\textwidth}
  \includegraphics[width=1.0\linewidth]{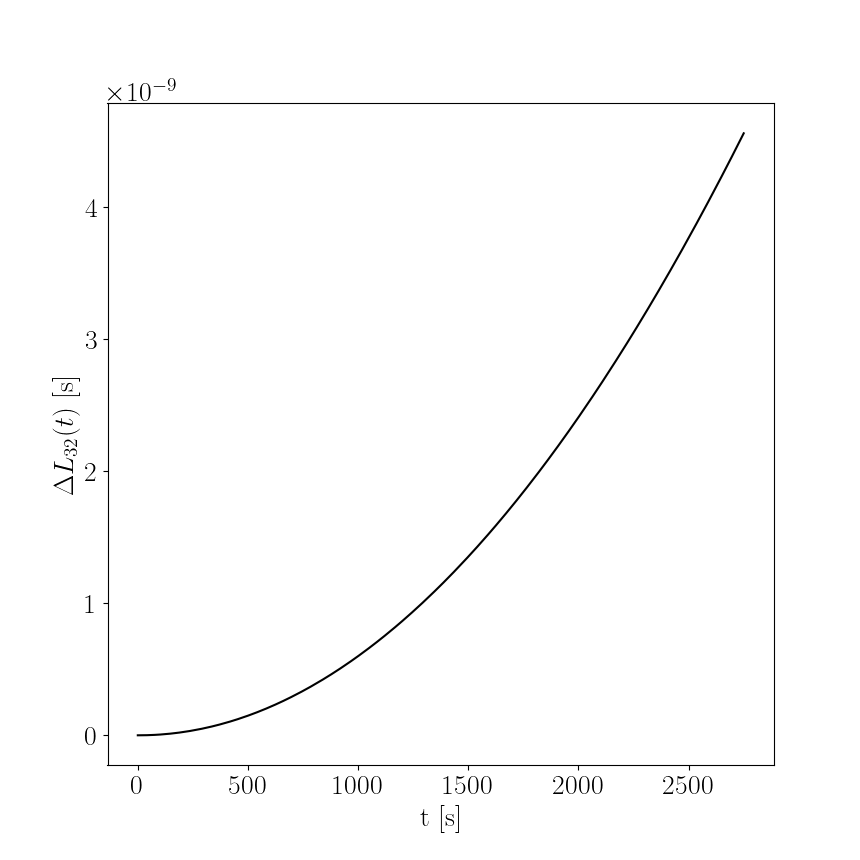}
  \caption{}
  \label{fig:Linear approximation_a}
\end{subfigure}%
\begin{subfigure}{.5\textwidth}
  \includegraphics[width=1.0\linewidth]{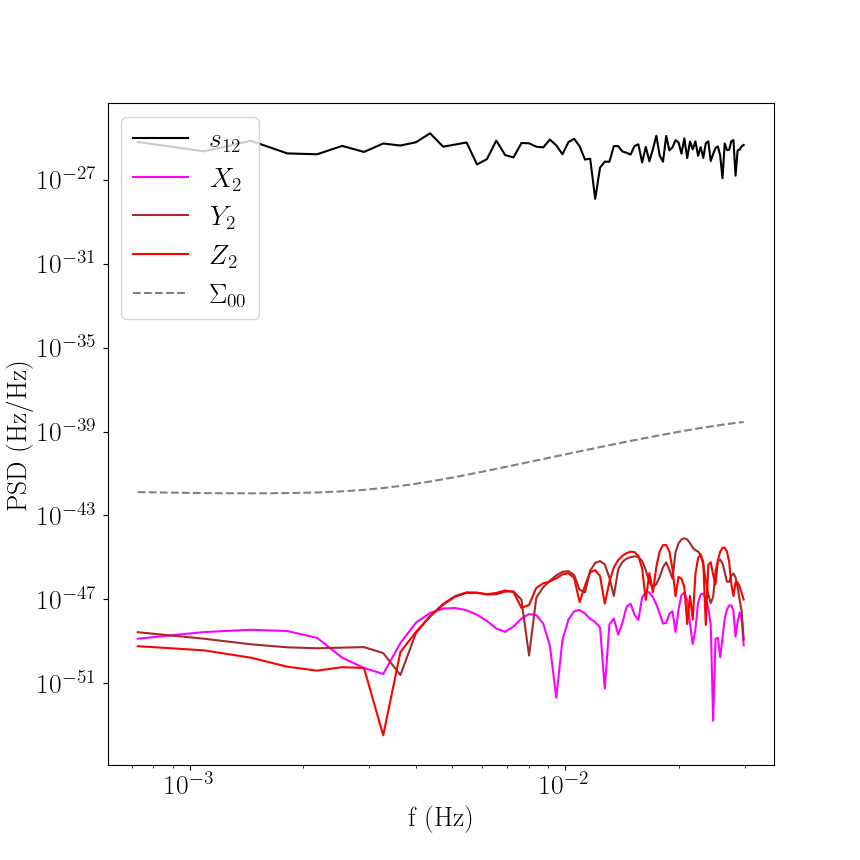}
  \caption{}
  \label{fig:Linear approximation_b}
\end{subfigure}
\caption{(a) Linear approximation - true time dependence ($\Delta L_{32}(t) = L^{\mathrm{LINEAR}}_{32}-L^{\mathrm{TRUTH}}_{32}$) for arm length $L_{32}(t)$ for LISA in Keplerian orbit on $\sim$ 1 hour of data. (b) TDI residuals $X_2(f)$ (magenta), $Y_2(f)$ (brown), $Z_2(f)$ (red) using linear approximation compared to the secondary noise reference curve (dashed grey) given by Eq. \ref{eq:diagonalCov}. Although the TDI residual suppresses LFN noise, the linear time dependence is not sufficient for estimating $L_{ij}$ posteriors.}
\label{fig:Linear approximation}
\end{figure*}

Since the second-generation combinations suppress LFN to first order in the spacecraft distance velocities, we analyzed a linear parameterization of the time-dependence by estimating all 6 $L_{ij}$ and 3 $\dot{L}_{ij}$ (since $\dot{L}_{ij} \approx \dot{L}_{ji}$). While we do see LFN suppression in this parameterization (Fig. \ref{fig:Linear approximation_b}), a comparison of the true $L_{ij}(t)$ in the Keplerian orbit model to the linear expansion is shown in Fig. \ref{fig:Linear approximation_a}. We additionally tried a quadratic approximation which added three more $\ddot{L}_{ij}$ parameters but all expansion-based $L_{ij}(t)$ parameterization methods result in biased posterior estimates, so a more accurate parameterization is required. 

\begin{figure*}
\begin{subfigure}{.5\linewidth}
  \includegraphics[width=1.0\linewidth]{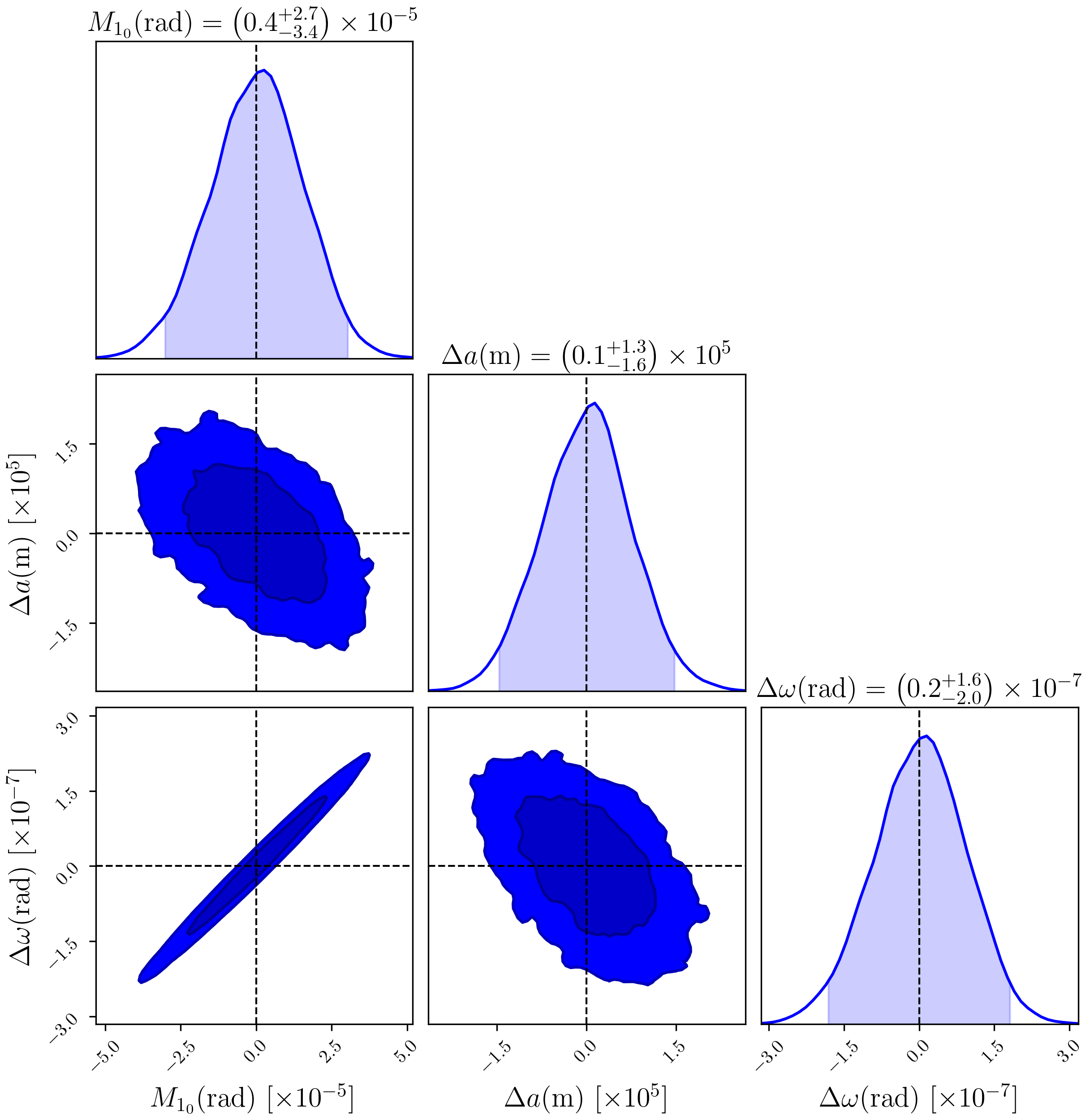}
  \caption{}
  \label{fig:Keplerian_results_a}
\end{subfigure}%
\begin{subfigure}{.5\linewidth}
  \includegraphics[width=1.0\linewidth]{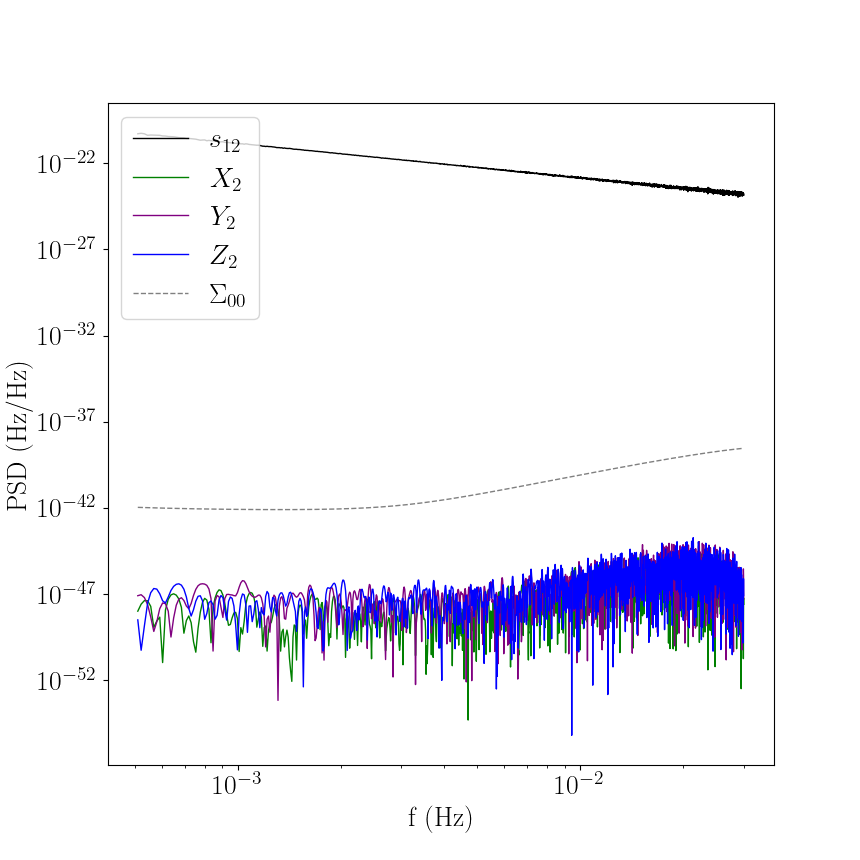}
  \caption{}
  \label{fig:Keplerian_results_b}
\end{subfigure}
\caption{(a) Posteriors for $\vec{\theta} = \{a,M_{0_1},\omega\}$ $\vec{L}_{ij}$ parameterization on Keplerian orbital data. The semi-major axis and argument of perihelion are shown as  differences from truth values used in the simulation ($\Delta a$ and $\Delta\omega$). 90\% credible interval quantities are quoted and $1\sigma$ (dark blue) and $2\sigma$ (blue) shaded contours are shown. (b) TDI residuals $X_2(f)$ (green), $Y_2(f)$ (purple), $Z_2(f)$ (blue) using median values from posterior distribution compared to the secondary noise reference curve (dashed grey) given by Eq. \ref{eq:diagonalCov}.}
\label{fig:Keplerian_results}
\end{figure*}

\section{Performance Testing}\label{results}

\subsection{Keplerian Parameterization}\label{KeplerianParameterization}

The light travel time $L_{ij}(t)$ is simply the time difference between reception on spacecraft $i$ from spacecraft $j$ and we follow the procedure written in the \verb|LISA Orbits| documentation based on a post-Minkowskian expansion given that the emission time is unknown upon reception which ultimately depends on the  positions (Eq. \ref{eq:fullKeplerPositions}), velocities and accelerations of the three spacecraft in the solar-system barycentric (SSB) reference frame at the time of reception. These quantities are derived from the orbital motion of each spacecraft using the 6 Keplerian orbital elements $\{a,e,\iota,M_0,\Omega,\omega\}$. 

\begin{alignat}{3}\label{eq:fullKeplerPositions}
    x_i(t) &= r_{c_i}(t)[
\cos{\nu_i(t)}(\cos{\omega}\cos{\Omega_i} - \sin{\omega}\cos{\iota}\sin{\Omega_i})\nonumber \\
&- \sin{\nu_i(t)}(\sin{\omega}\cos{\Omega_i} + \cos{\omega}\cos{\iota}\sin{\Omega_i})]\nonumber  \\
    y_i(t) &= r_{c_i}(t)[
\cos{\nu_i(t)}(\cos{\omega}\sin{\Omega_i} + \sin{\omega}\cos{\iota}\cos{\Omega_i})\nonumber \\
&+ \sin{\nu_i(t)}(\cos{\omega}\cos{\iota}\cos{\Omega_i} - \sin{\omega}\sin{\Omega_i})]\nonumber \\
z_i(t) &=r_{c_i}(t)[\cos{\nu_i(t)}\sin{\omega}\sin{\iota} + \sin{\nu_i(t)}\cos{\omega}\sin{\iota}]
\end{alignat}
where $r_{c_i}(t) = a(1-e\cos{E_i(t)})$ is the distance to the solar system barycenter and $E_i(t)$ is the eccentric anomaly related to the true anomaly $\nu_i(t)$ by $\nu_i(t) = \arctan{\frac{\sqrt{1+e}}{\sqrt{1-e}}\tan{\frac{E_i(t)}{2}}}$. The velocities are given by taking the analytical derivative of Eq. \ref{eq:fullKeplerPositions}, and the accelerations $\ddot{r}_i$ are $$\ddot{r}_i = \mu_{\odot}\frac{\vec{r}_i}{|\vec{r}_i|^3},$$
where $\mu_{\odot}$ is the solar gravitational parameter $\mu_{\odot} = GM_{\odot}$. 

In the case of optimized Keplerian orbits for LISA, all 3 spacecraft share the same argument of perihelion $\omega$ and semi-major axis $a$ which defines the eccentricity $e$ and inclination $\iota$ given a tilt angle factor of $\delta = 5/8$ to minimize the constellation breathing \cite{delta58}. The eccentric anomaly $E_(t)$ is related to the initial mean anomaly parameter $M_0$ by Kepler's equation $M(t) = M_0 + (t-t_0)\frac{\mu_{\odot}}{a^3} = E(t) -e\sin{E(t)}$ and $M_{i_0} = M_{1_0} - 2(i-1)\pi/3$ for spacecraft $i \in \{1,2,3\}$. The longitude of the ascending node for spacecraft $i$ is $\Omega_i = \Omega_1 + 2(i-1)\pi/3$. The longitude of the ascending node is unconstrained due to the LISA orbit. When the inclination $\iota=0$, $\Omega$ is undefined, and our orbital elements are described in ecliptic coordinates in the barycentric reference frame. The LISA guiding center will be in an Earth-leading or Earth-trailing heliocentric orbit, making the inclination $i\sim 0$. Therefore the only parameters required to describe all 6 armlengths $L_{ij}(t)$ in the Keplerian model are $$\vec{\theta} = \{a,M_{0_1},\omega\} \longrightarrow {L_{ij}(t)}.$$ 

We use the \verb\emcee\ affine-invariant MCMC ensemble sampler \cite{emcee} on a 1-day duration of data simulated with the Keplerian orbit model using the ``stretch move'' proposal jumps. We use a uniform prior over each parameter with $p(a) = \mathcal{U}[\SI{1}{AU} - \SI{2e8}{m},\SI{1}{AU}+\SI{1e8}{m}]$ which spans the linear range over a 10 year LISA mission as estimated in the numerical model given in \cite{Trajectory_Design} centered at $\SI{1}{AU}$. The remaining angular orbital element priors are $p(M_{0_1}) = \mathcal{U}[-\pi,\pi]$ to avoid incomplete distributions at the prior boundaries and $p(\omega) = \mathcal{U}[-\pi,0]$ under the assumption of clockwise rotation of the spacecraft. To reduce computational cost even further for a day's worth of data with a time-varying FDI filter, we reduce the sampling rate to 2 Hz and set the maximum frequency in the likelihood function to $f_{\mathrm{max}} = 0.03$ Hz since we are primarily concerned with LFN suppression while maintaining a significant portion of the LISA signal band in the analysis. The new FDI filter (Sec. \ref{newFilterSection}) to delay the measurements for each MCMC iteration now requires a filter length of only $N=7$ which saves on run-time and significantly reduces the amount of data loss.

Figure \ref{fig:Keplerian_results_a} shows the $\Delta a, M_{0_1}$ and $\Delta \omega$ posterior distributions and the 90\% credible intervals centered around the truth values used in the data simulation. The semi-major axis and the initial mean anomaly of spacecraft 1 are represented by differences from their truth values that were input into the data simulation for figure clarity. The $1\sigma$ (dark blue) and $2\sigma$ (blue) shaded contours are shown. Figure \ref{fig:Keplerian_results_b} show the the TDI residuals using the median posterior estimates for $\{a,M_{0_1},\omega\}$ compared to one of the science measurements containing LFN to demonstrate sufficient noise suppression. The boundaries of the 90\% credible intervals also produce TDI residuals well below the secondary noises.

\subsection{A Proposed Keplerian Model For Realistic Numerical LISA Orbits}\label{arbitraryOrbits}

\begin{figure}
    \includegraphics[width=1.0\linewidth]{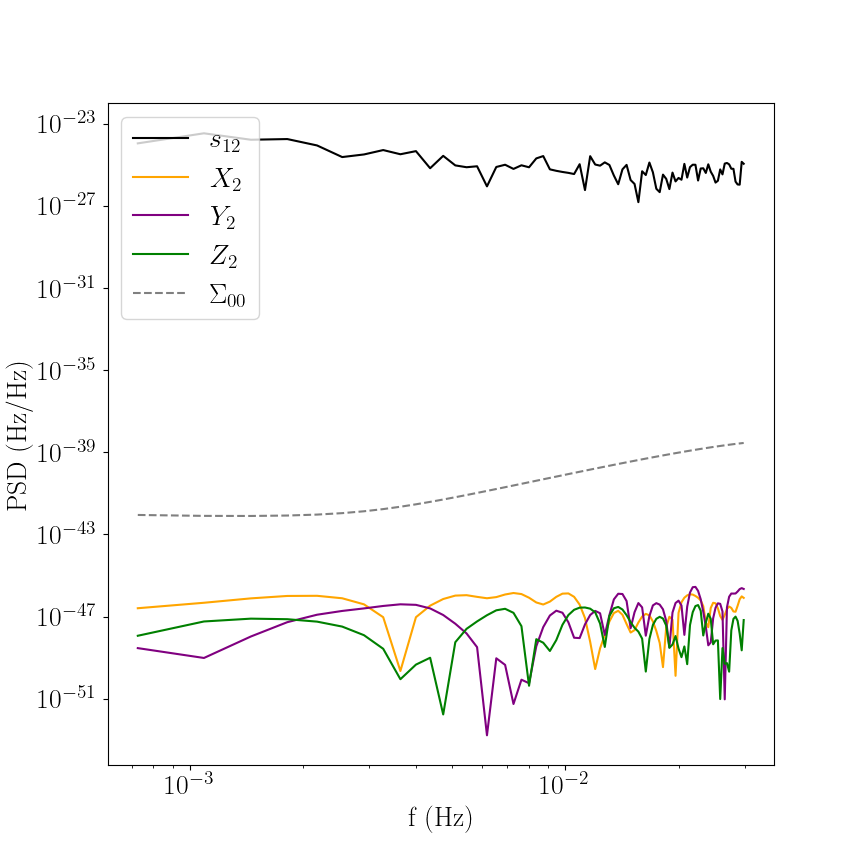}
    \caption{TDI residuals $X_2(f)$ (orange), $Y_2(f)$ (purple), $Z_2(f)$ (green) on 1 hour of arbitrary numerical orbit data using constant-valued Keplerian orbital elements compared to the secondary noise reference curve (dashed grey) given by Eq. \ref{eq:diagonalCov}.}
    \label{fig:one_hour_elements_from_Cartesian}
\end{figure}

\begin{figure*}
\begin{subfigure}{.5\textwidth}
  \includegraphics[width=1.0\linewidth]{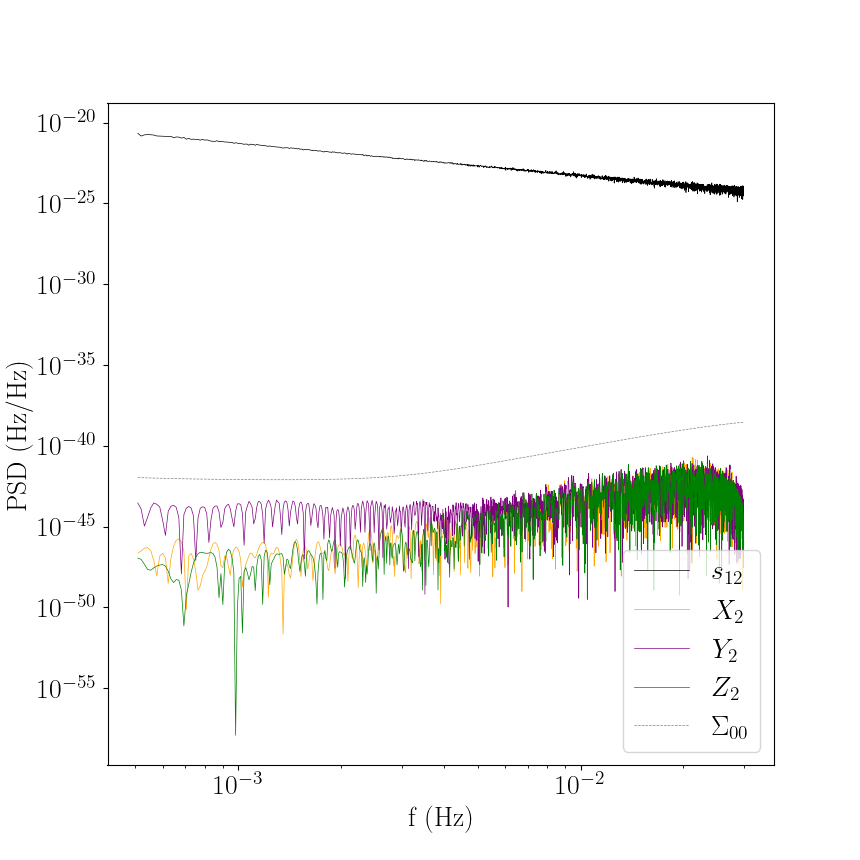}
  \caption{}
  \label{fig:ESA_1_day_residuals_a}
\end{subfigure}%
\begin{subfigure}{.5\textwidth}
  \includegraphics[width=1.0\linewidth]{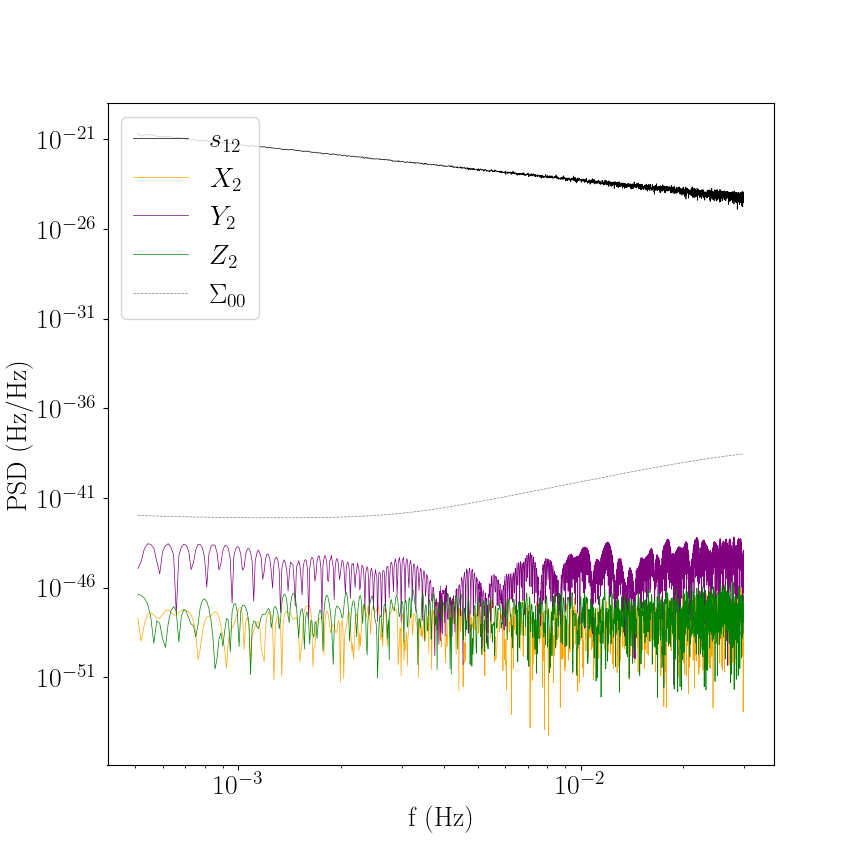}
  \caption{}
  \label{fig:ESA_1_day_residuals_b}
\end{subfigure}
\caption{TDI residuals $X_2(f)$ (orange), $Y_2(f)$ (purple), $Z_2(f)$ (green) on 1 day of arbitrary numerical orbit data using constant-valued Keplerian orbital elements compared to the secondary noise reference curve (dashed grey) given by Eq. \ref{eq:diagonalCov} for: (a) True Positions and Velocities from orbit data simulation. (b) Initial values in time for the Keplerian orbital elements of each S/C.} 
\label{fig:ESA_1_day_residuals}
\end{figure*}

The more realistic orbit model used in the data simulation accounts for the other forces that act upon the LISA spacecraft such as the gravitational force of other solar system bodies, especially the Earth's. The numerical model described in \cite{Trajectory_Design} and used in \verb|LISA Orbits| points out that the Earth's gravity leads to a linear drift in the semi-major axis and is the most significant perturbation to analytical Keplerian orbits. We tested whether our sampler could estimate the time-varying spacecraft separations using an adiabatic Keplerian model, fitting Keplerian orbital elements to the spacecraft orbits on short time segments. Now the spacecraft motion cannot be described by 6 orbital elements, but rather an osculating orbit with time-varying orbital parameters for all three spacecraft. If we can sample the orbital elements on a sufficiently short duration of data such that the time-variance of the orbital parameters is negligible for calculation of $\vec{L}_{ij}$ and sufficient LFN suppression, the posterior results can be interpolated across a chosen duration of LISA data to estimate the time-varying Keplerian orbital elements on any chosen scale.

We began by looking at the position and velocity for each spacecraft in the SSB time frame for 1 hour of numerical data and converted them to time-varying Keplerian orbital elements. Figure \ref{fig:one_hour_elements_from_Cartesian} demonstrates adequate LFN suppression using the first time sample of the converted orbital elements as input to the position equations in Eq. \ref{eq:fullKeplerPositions}, now with an ``i'' subscript after every orbital element since they all differ for each spacecraft. Use of the first sample for the $M_{0_i}$ parameter is crucial for accurate spacecraft separations, but one could explore using median values or other constant valued representations for the other parameters. We then examined sampling this 18-parameter model with $$\vec{\theta}_i = \{a_i,e_i,\iota_i,M_{0_i},\omega_i,\Omega_i\},  i \in \{1,2,3\}$$ for each spacecraft $i$ on the hour-long data segment using MCMC methods since constant values of the time-varying parameters suppress LFN. This is a challenging parameter model to sample, and work is still in progress to find the best sampling techniques to gain efficiently processed posteriors. We noticed it performed exceptionally better when the uniform prior ranges for the semi-major axis were narrowed from the 10-year expected range to a smaller range indicating they may need adjustment depending on the start time and duration of the segment. It is a realistic expectation that some mission data will provide more informative priors (e.g. narrowed uniform distributions) to speed up sampling convergence.

Figure \ref{fig:ESA_1_day_residuals} compares the TDI residuals using the constant-valued orbital elements to calculate the arm lengths versus the true positions and velocities from the orbit simulation used as input to the data simulation for a 1-day duration. Constant values still suppress LFN, but one needs to examine whether the resulting $\log{\mathcal{L}}$ values are comparable to the true positions for accurate posterior distributions. For a sufficiently sampled posterior chain, the $\Delta\log{\mathcal{L}} \propto N_D/2$ where $N_D$ is the model dimension. Therefore we would want the $\log{\mathcal{L}}$ difference to be $\lessapprox$ 9 for an 18-parameter Keplerian model. We find that a duration of 9 hours of numerical LISA data sampled at 4 Hz gives a $\Delta\log{\mathcal{L}} = 7.4$ using a filter length $N=7$ and maximum frequency $f_{\mathrm{max}} = 0.03$ Hz. One could sample 9 hour segments at a time of LISA data and interpolate point estimates and credible intervals of the resulting posterior distribution to estimate the LISA orbits independent of any mission ephemeris data as a back-up and compliment to on-board phase modulating psuedo-ranging once future work demonstrates a successful Bayesian sampling of the 18 parameter Keplerian model.

\section{Conclusion And Future Work}

The goal of Bayesian data-driven methods for estimating the TDI delays is acquiring posterior estimates on the arm length parameters to be marginalized over when computing the LISA response function to GWs. We demonstrate the challenges and ways to mitigate them in the realistic LISA scenario in heliocentric orbit. We use a Taylor-expanded version of the FDI filter to compute and store approximately half of the time-varying filter operations prior to running MCMC iterations to significantly improve computational efficiency. The time-dependent delays are estimated using Keplerian orbital elements for the Keplerian orbit model and provide a foundation for this parameterization in the realistic arbitrary orbit scenario. Future efforts to build upon this data driven TDI strategy will move forward with MCMC sampling algorithms to fully characterize the LISA spacecraft orbits by individual estimation of approximately 9 hour segments of data using a $\SI{4}{Hz}$ sampling rate and interpolating segment posteriors. Arm length posteriors can then be incorporated into the LISA response function to GWs, and marginalization over arm-length uncertainty will remove potential sources of bias during GW signal parameter estimation.

\section{Acknowledgements}

This project was supported by the NASA LISA Study Office. 

\bibliography{references}

\end{document}